\documentclass[journal,twoside,web]{ieeecolor}
\usepackage{generic}
\usepackage{cite}
\usepackage{amsmath,amssymb,amsfonts}
\usepackage{algorithmic}
\usepackage{graphicx}
\usepackage{textcomp}
\usepackage{setspace}
\usepackage{makecell}
\usepackage{textcomp}
\def\BibTeX{{\rm B\kern-.05em{\sc i\kern-.025em b}\kern-.08em
    T\kern-.1667em\lower.7ex\hbox{E}\kern-.125emX}}
\begin{document}
\title{Characterization and Modeling of Native MOSFETs Down to 4.2 K}
\author{Yuanke Zhang, Tengteng Lu, Wenjie Wang, Yujing Zhang, Jun Xu, Chao Luo, Guoping Guo
\thanks{This work was supported by the National Natural Science Foundation of China (Grants No. 12034018). (Corresponding author: Chao Luo, e-mail: lc0121@ustc.edu.cn)}    
\thanks{The authors are with Key Laboratory of Quantum Information, University of Science and Technology of China, Hefei, Anhui 230026, China.}    
}
\maketitle
\begin{abstract}
The extremely low threshold voltage ($V_{TH}$) of native MOSFETs ($V_{TH}$$\approx$0$\,$V$\,$@$\,$300$\,$K) is conducive to the design of cryogenic circuits. Previous research on cryogenic MOSFETs mainly focused on the standard threshold voltage (SVT) and low threshold voltage (LVT) MOSFETs. In this paper, we characterize native MOSFETs  within the temperature range from 300$\,$K to 4.2$\,$K. The cryogenic V$_{TH}$ increases up to $\sim$0.25$\,$V (W/L$\,$=$\,$10$\,$$\mu$m/10$\,$$\mu$m) and the improved subthreshold swing (SS)$\,$$\approx$$\,$14.30$\,$mV/dec$\,$@$\,$4.2$\,$K. The off-state current ($I_{OFF}$) and the gate-induced drain leakage (GIDL) effect are ameliorated greatly. The step-up effect caused by the substrate charge and the transconductance peak effect caused by the energy quantization in different sub-bands are also discussed. Based on the EKV model, we modified the mobility calculation equations and proposed a compact model of large size native MOSFETs suitable for the range of 300$\,$K to 4.2$\,$K. The mobility-related parameters are extracted via a machine learning approach and the temperature dependences of the scattering mechanisms are analyzed. This work is beneficial to both the research on cryogenic MOSFETs modeling and the design of cryogenic CMOS circuits for quantum chips.
\end{abstract}


\section{Introduction}
\label{sec:introduction}
\IEEEPARstart{C}{ryogenic} CMOS (cryo-CMOS) technology has been studied widely in recent years and plays an important role in the readout and control circuits used for quantum chips\cite{b1}\cite{b2}. Cryogenic operation of CMOS circuits improves both the scalability and the system integration of quantum chips. Meanwhile, the quantum effects of cryogenic MOSFETs (e.g., Coulomb blocking) enable cryo-CMOS to build not only classical circuits but also quantum circuits\cite{b3}. However, cryo-CMOS technology still faces several challenges, including power limitations and cryogenic MOSFETs modeling. Because of the huge temperature gap and low-temperature effects involved, most of the classic models (e.g., BSIM3v3\cite{b8}, EKV2.6\cite{b19}) are only suitable around room temperature (RT). Therefore, the research and modeling of cryogenic MOSFETs are necessary.
\par To date, characterization of MOSFETs technology in the range of 0.35$\,$$\mu$m to 14$\,$nm from 77$\,$K to 50$\,$mK has been widely reported\cite{b3,b4,b5,b6,b7,b21}. Most of these studies have focused on standard threshold voltage (SVT) and low threshold voltage (LVT) MOSFETs, but few have characterized native MOSFETs ($V_{TH}$$\approx$0$\,$V$\,$@$\,$300$\,$K), which can be used in high-speed and low operating voltage circuits. Because of the increase of substrate Fermi potential at cryogenic temperature, the high threshold voltages of cryogenic MOSFETs increase the design difficulty. Native MOSFETs represent a good choice ($V_{TH}$$<$0.3$\,$V$\,$@$\,$4.2$\,$K). In addition, the kink effect\cite{b6} is alleviated in native MOSFETs because of their low substrate doping concentration ($Na$) and low body effect parameter ($\gamma=\sqrt{2q\varepsilon_{s i}Na}/C_{OX}$, $C_{OX}$ is the gate oxide capacitance per unit area), thus simplifying the device modeling process.
\par In this paper, the low-temperature characteristics of native MOSFETs at temperatures ranging from 300$\,$K to 4.2$\,$K are presented, with particular focus on the step-up effect and the transconductance peak effect. In addition, the current drop (i.e., negative transconductance) behavior in the linear region transfer characteristics is the most noteworthy. By modifying the mobility equations of the EKV2.6 model, the mobility is calculated using scattering mechanisms, i.e., Coulomb scattering, surface roughness scattering, and phonon scattering, and the current drop behavior is reproduced well. Furthermore, the mobility-related parameters are extracted using a machine learning approach, and the temperature dependences of the scattering mechanisms are analyzed. Unlike the standard EKV2.6 model with its minimum applicable temperature of 230$\,$K, the compact model of the large size native MOSFETs proposed in this paper is suitable for an extended temperature range from 300$\,$K down to 4.2$\,$K.
\section{Cryogenic Measurement Setup}
The devices under test (DUT) in this paper were manufactured in a commercial SMIC 0.18$\,$$\mu$m bulk MOSFETs process with 1.8$\,$V nominal voltage. DUT were n-type native MOSFETs with W/L$\,$=$\,$10$\,$$\mu$m/10$\,$$\mu$m and 10$\,$$\mu$m/1$\,$$\mu$m. We also characterized some SVT and LVT MOSFETs as the comparison. DC characteristics were performed by a Keysight B1500A semiconductor device analyzer, and the 4-wire method was taken to remove the influence of wire resistance. Liquid nitrogen/helium dewar was used to cool DUT from 300$\,$K to 77$\,$K/4.2$\,$K. Transfer characteristics in linear/saturation regions ($I_{DS}$-$V_{GS}$, $V_{DS}$$\,$=$\,$50$\,$mV/1.8$\,$V) and output characteristics ($I_{DS}$-$V_{DS}$, $V_{GS}$$\,$=$\,$0$\,$V$\rightarrow$1.8$\,$V, step$\,$=$\,$0.3$\,$V) were measured. The electrical properties of DUT were extracted: $V_{TH}$ (using maximum transconductance method\cite{b3}), the drain saturation current ($I_{DSAT}$, $V_{DS}$$\,$=$\,$$V_{GS}$$\,$=$\,$1.8$\,$V), the OFF-state drain leakage current ($I_{OFF}$, $V_{DS}$$\,$=$\,$1.8$\,$V and $V_{GS}$$\,$=$\,$0$\,$V), and the drain induced barrier lowering (DIBL) effect which is defined as [$V_{GS}$($I_{DS}$\,$=$\,$5\times10^{-7} A, V_{DS}$\,$=$\,$50$$\,$mV)$\,$-$\,$V$_{GS}$($I_{DS}$\,$=$\,$5\times10^{-7} A, V_{DS}=1.8$$\,$V)]/1.75$\,$V. 
\section{Characterization}
\subsection{Cryogenic Measurement}
\begin{figure}[!h]
	\centerline{\includegraphics[width=\columnwidth]{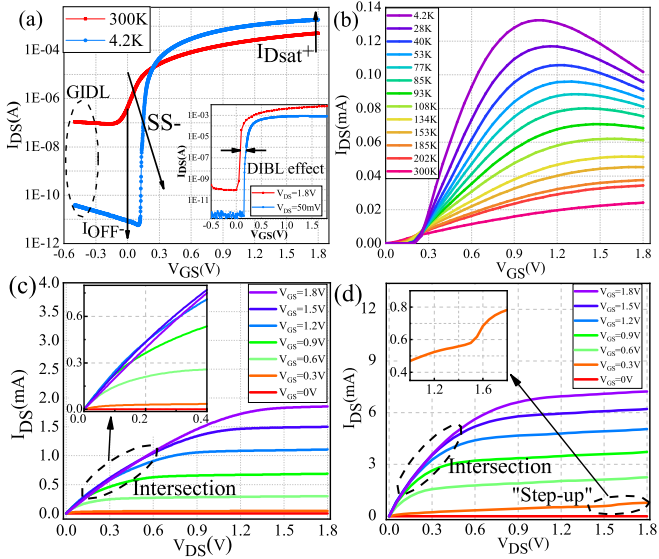}}
	\caption{DC Characteristics of native MOSFETs. (a) Comparison of the transfer characteristics in the saturation regions ($V_{DS}$$\,$=$\,$1.8$\,$V, W/L$\,$=$\,$10 $\mu$m/10 $\mu$m) at 300$\,$K and 4.2$\,$K, respectively. Inset: transfer characteristics in the linear region and the saturation region ($V_{DS}$$\,$=$\,$50$\,$mV/1.8$\,$V, W/L$\,$=$\,$10$\,$$\mu$m/1$\,$$\mu$m) at 4.2$\,$K. (b) Transfer characteristics in the linear region at various temperatures.(c) Output characteristics with W/L$\,$=$\,$10$\,$$\mu$m/10$\,$$\mu$m at 4.2$\,$K. Inset: the enlarged view of the part circled by the dotted line (the intersection of output characteristics curves). (d) Output characteristics with W/L$\,$=$\,$10 $\mu$m/1 $\mu$m at 4.2$\,$K. Inset: the enlarged view of the part circled by the dotted line (the step-up effect). }
	\label{fig1}
\end{figure}

Cryogenic operation can greatly improve MOSFETs performance. As shown in Fig. \ref{fig1}(a), we extracted $V_{TH}$, subthreshold swing ($SS$), $I_{DSAT}$, and $I_{OFF}$ from the transfer characteristics at 300$\,$K and 4.2$\,$K, and listed them in Table \ref{tab1}. At low temperatures, the intrinsic carrier concentration decreases and the Fermi energy level of the p-type substrate moves closer to the valence band, thus resulting in a significant increase of the Fermi potential. Therefore, $V_{TH}$ increases  to $\sim$0.25$\,$V. $SS$ improves significantly, but is still much higher than the theoretical limit ($ln(10)K_{B}T/q$, $\sim$$0.83\,mV/dec$$\,$@$\,$4.2$\,$K); this can be attributed to the effect of the interface traps close to the band edges\cite{b5}. Additionally, the reduced $I_{OFF}$ and improved $I_{DSAT}$ stronglyly optimize the device ON/OFF ratio. The DIBL effect and the gate-induced drain leakage (GIDL) effect are also observed, as shown in Fig. \ref{fig1}(a). The reduced ionization rate causes the source/drain p-n$^{+}$ junction depletion to broaden, thus reducing the effective channel length and aggravating the DIBL effect. The GIDL effect can be explained using the band-to-band tunneling (BTBT) process. At cryogenic temperatures, the larger bandgap elongates the tunneling distance, which then weakens the BTBT effect and reduces the leakage current in the negative $V_{GS}$ region\cite{b9}. 
\par Fig. \ref{fig1}(b) shows the linear region transfer characteristics from 300$\,$K to 4.2$\,$K. It should be noted that, with decreasing temperature, the curves show more pronounced current drop (i.e., negative transconductance) behaviors. This is causd by the change of the mobility behavior: at low temperatures, when $V_{GS}$ increases, the increase of the number of carriers cannot compensate for the reduction of carrier mobility, thus causing $I_{DS}$ to decrease. Fig. \ref{fig1}(c) and (d) show the output characteristics of native MOSFETs (W/L$\,$=$\,$10$\,$$\mu$m/10$\,$$\mu$m and 10$\,$$\mu$m/1$\,$$\mu$m). Intersection of these curves can be observed in both figures. This phenomenon corresponds to the linear region current drop behavior that was described above: different $V_{GS}$ may results in the same $I_{DS}$ under the same $V_{DS}$. This phenomenon is described and discussed in detail in Section III. In addition, in Fig. \ref{fig1}(d), a step-up effect of a sudden $I_{DS}$ increase is observed. To explore the physical mechanism behind this effect, we performed measurements with different delay times\cite{b7} at specific temperatures, with results as shown in Fig. \ref{fig2}.

\begin{table}
	\caption{DC CHARACTERISTICS OF NATIVE NMOSFETS (W/L = 10 $\mu m$/10 $\mu m$) AT 300 K AND 4.2 K}
	\centering
	\label{table}
	\setlength{\tabcolsep}{6pt}
	\begin{spacing}{1}
	\begin{tabular}{|m{3.7cm}<{\centering}|m{1.8cm}<{\centering}|m{1.8cm}<{\centering}|}
		\hline
		Temperature& 
		300 K& 
		4.2 K\\
		\hline
		$V_{TH} (mV)$& 
		38.05& 
		257.07 \\
		\hline
		$SS (mV/dec)$& 
		86.25& 
		14.30 \\
		\hline
		$I_{DSAT} (mA)$& 
		0.511& 
		1.852 \\
		\hline
		$I_{OFF} (A)$& 
		$4.39\times10^{-7}$& 
		$8.70\times10^{-12}$ \\
		\hline
		$I_{DSAT}/I_{OFF}$& 
		$1.16\times10^{3}$& 
		$2.13\times10^{8}$ \\
		\hline
		$DIBL (mV/V)$  (W/L = 10 $\mu m$/1 $\mu m$)& 
		35.429& 
		80.571 \\
		\hline
	\end{tabular}
    \end{spacing}
	\label{tab1}
\end{table}

\subsection{Step-Up Effect}
\begin{figure*}[!h]
	\centerline{\includegraphics[scale=1]{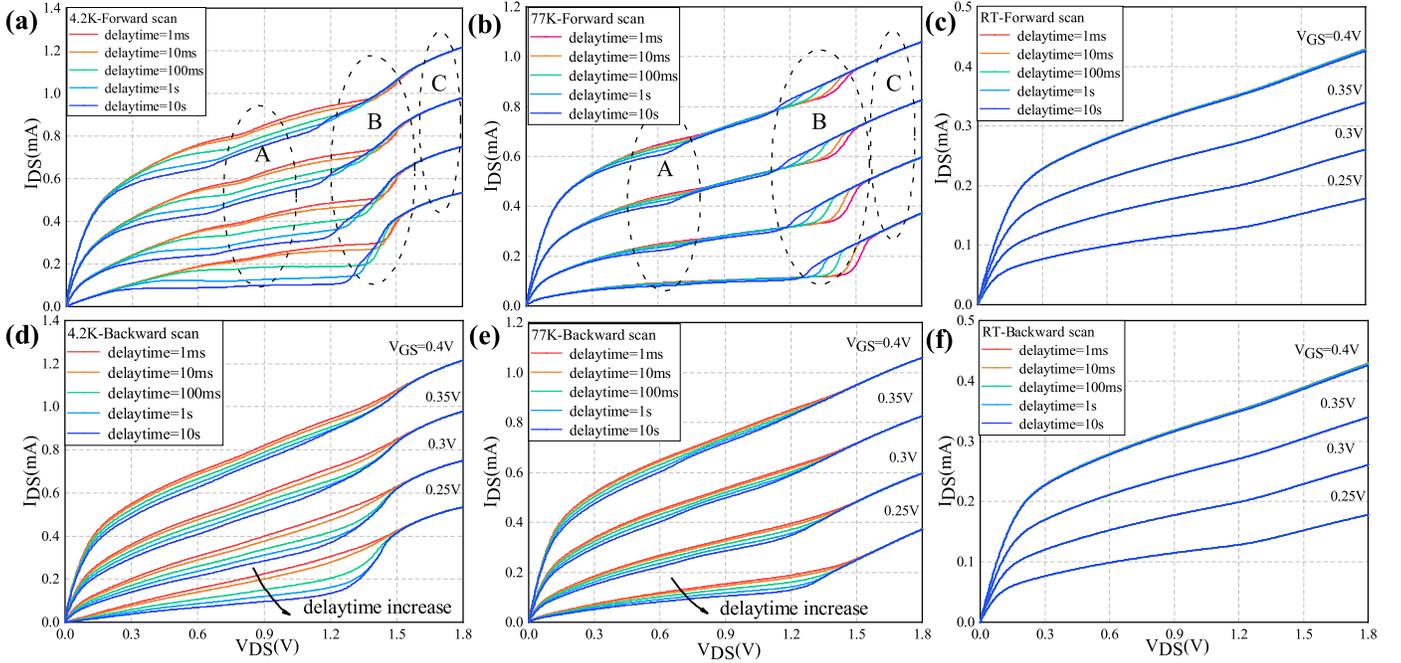}}
	\caption{Forward scans (a-c) and backward scans (d-f) of I-V characteristics for the native MOSFETs with W/L$\,$=$\,$10$\,$$\mu m$/1$\,$$\mu m$ at different temperatures. (a,d) At 4.2$\,$K. (b,e) At$\,$77 K. (c,f) At RT. $V_{GS}$$\,$=$\,$0.25$\,$V/0.3$\,$V/0.35$\,$V/0.4$\,$V.}
	\label{fig2}
\end{figure*}
We applied delay times ranging from 1$\,$ms to 10$\,$s and performed both forward scans [Figs. \ref{fig2}(a)-(c)] and backward scans [Figs. \ref{fig2}(d)-(f)]. The splitting and the step-up behaviors of $I_{DS}$ curves were observed under different delay times. Two mechanisms are involved in the above phenomena: the accumulation of substrate holes due to impact ionization and the dissipation of these holes via substrate/source parasitic capacitance, and both mechanisms are time-dependent. At low temperatures, the holes generated by impact ionization flow towards and accumulate in the freeze-out substrate, thus raising the substrate potential; this leads to a reduction of $V_{TH}$ and thus a rise of $I_{DS}$, which is similar to the kink effect that occurs in SVT MOSFETs\cite{b6}. Additionally, the accumulated holes in the substrate can also be transferred to the source via the substrate/source parasitic capacitance, thus reducing the substrate potential.
\par In the lower $V_{DS}$ range, e.g., region A in Fig. \ref{fig2}(a) and (b), a higher delay time results in a reduced $I_{DS}$. This is because carriers impact ionization produces fewer holes at lower $V_{DS}$. A longer delay time results in more complete hole dissipation in the substrate, i.e., lower substrate potential and thus smaller$\,$$\,$$I_{DS}$. 
\par In the regions where $I_{DS}$ rise occurs, i.e., region B in Fig. \ref{fig2}(a) and (b), because of the higher $V_{DS}$, the number of holes generated by impact ionization is greater than that dissipated by the parasitic capacitor, resulting in a rapid rise of $I_{DS}$. Therefore, the higher delay time causes more holes to accumulate in the substrate, and thus $I_{DS}$ under the higher delay time becomes larger than that under the lower delay time. When enough holes accumulate in the substrate,  the substrate potential is so high that the substrate/source p-n$^{+}$ junction is forward-biased; this results in an almost constant substrate potential and $I_{DS}$ tends to be saturated\cite{b6}. Therefore, when $V_{DS}$ is sufficiently large, even under different delay times, there are enough holes accumulating in the substrate, resulting in the same substrate potential and the same $I_ {DS}$, as shown in region C in Fig. \ref{fig2}(a) and (b). 
\par
\begin{figure}[!t]
	\centerline{\includegraphics[width=\columnwidth]{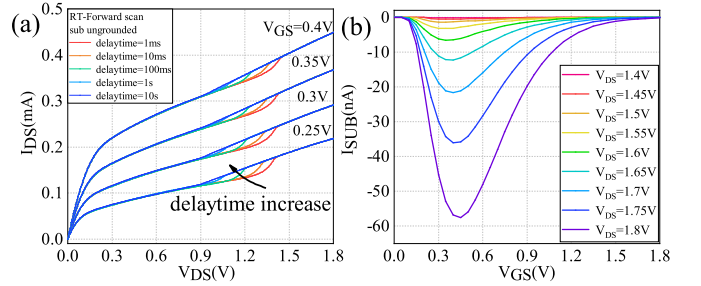}}
	\caption{For the native MOSFETs with W/L$\,$=$\,$10$\,$$\mu m$/1$\,$$\mu m$, (a) I-V characteristics when the substrate electrode is ungrounded at room temperature. (b) Substrate current($I_{SUB}$) at 4.2$\,$K.}
	\label{fig3}
\end{figure}

 At room temperature, the holes can flow to the substrate electrode without accumulation, hence no step-up  phenomenon is observed, as shown in Fig. \ref{fig2}(c) and (f). To provide further verification that this phenomenon is related to the substrate potential, we set the substrate electrode to be ungrounded at room temperature, and the step-up effect appears as expected, as shown in Fig. \ref{fig3}(a). Furthermore, the measurement of the substrate current shown in Fig. \ref{fig3}(b) indicates that the impact ionization is more intense at lower $V_{GS}$. Therefore, the step-up effect is more pronounced at lower $V_{GS}$. 
\par When $V_{DS}$ is swept in the high-to-low direction, sufficient numbers of holes can accumulate in the substrate instantly because of the high initial $V_{DS}$. The substrate potential remains constant under higher $V_{GS}$ and decreases monotonously under lower $V_{GS}$. Therefore, only the splitting behaviors of the curves are observed, as shown in Fig. \ref{fig2}(d) and (e). In addition, at lower temperatures, the reduced ionization rate leads to a wider substrate/source depletion width\cite{b9}. The capacitance and the time constant of the parasitic capacitor are reduced and thus more holes move to the source through the parasitic capacitance per unit time. Therefore, the splitting behavior becomes more obvious at lower temperatures.
\subsection{Sub-Band Behavior}
\begin{figure}[!t]
	\centerline{\includegraphics[width=\columnwidth]{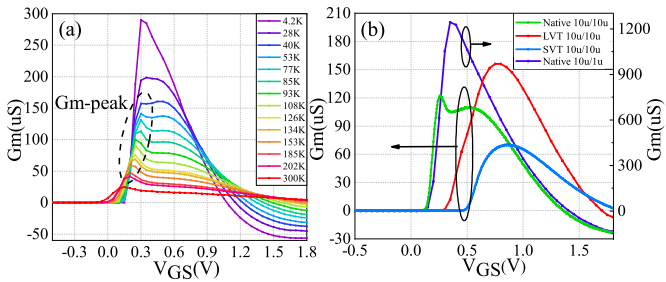}}
	\caption{(a) Transconductance characteristics of native MOSFETs at various temperatures. $V_{DS}$$\,$=$\,$50$\,$mV, W/L$\,$=$\,$10$\,$$\mu m$/10$\,$$\mu m$. (b) Comparison of transconductance characteristics of SVT MOSFET (W/L$\,$=$\,$10$\,$$\mu m$/10$\,$$\mu m$), LVT MOSFET (W/L$\,$=$\,$10$\,$$\mu m$/10$\,$$\mu m$), and native MOSFETs (W/L$\,$=$\,$10$\,$$\mu m$/10$\,$$\mu m$ and W/L$\,$=$\,$10$\,$$\mu m$/1$\,$$\mu m$) at 77$\,$K.}
	\label{fig4}
\end{figure}

As shown in Fig. \ref{fig4}(a), a peak transconductance value ($G_{m}$-peak) is observed from 40$\,$K to 202$\,$K. This peak value can be attributed to the energy quantization of different sub-bands in the inversion layer\cite{b22}. At intermediate temperatures, the carriers occupy the lowest and some higher energy sub-bands with different mobility. $V_{GS}$ changes the distribution of electrons in different sub-bands, thus resulting in a sudden change of the transconductance. However, at lower temperatures, only the lowest energy sub-band is filled and thus no $G_{m}$-peak behavior occurs below 40$\,$K. The occupancy of different sub-bands is related to the longitudinal (i.e., perpendicular to the channel) effective electric field ($E_{eff}$), which can be calculated as\cite{b12}:
\begin{equation}
E_{eff}=\left(\eta q_{inv}+q_{dep}\right) / \varepsilon_{s}
\end{equation}
where $\eta=0.5$ at room temperature and increases at cryogenic temperature. $q_{inv}$ is the inversion charge per unit area, $q_{dep}$ is the depletion charge per unit area and can be given by:
\begin{equation}
 q_{dep}=\sqrt{2q\varepsilon_{s i}Na\phi_{s}}
\end{equation}
where $\phi_{s}$ is the surface potential and $Na$ is the substrate doping concentration. To generate the $G_{m}$-peak behavior, $E_{eff}$ must be sufficiently low\cite{b22}, which requires small $q_{dep}$. The extremely low substrate doping of native MOSFETs can meet this requirement and thus no $G_{m}$-peak behavior is observed in the LVT (medium $Na$) and SVT (standard $Na$) MOSFETs at 77$\,$K, as shown in Fig. \ref{fig4}(b). Because of the reverse short channel effect (RSCE), the effective $Na$ increases with  decreasing channel length. Therefore, $G_{m}$-peak behavior is not observed in the native MOSFET with W/L$\,$=$\,$10$\,$$\mu$m/1$\,$$\mu$m, which is consistent with the opinion expressed in \cite{b11}.
 
\section{Cryo-MOSFETs Modeling}
\subsection{Method: Machine Learning}
\begin{figure}[!t]
	\centerline{\includegraphics[width=\columnwidth]{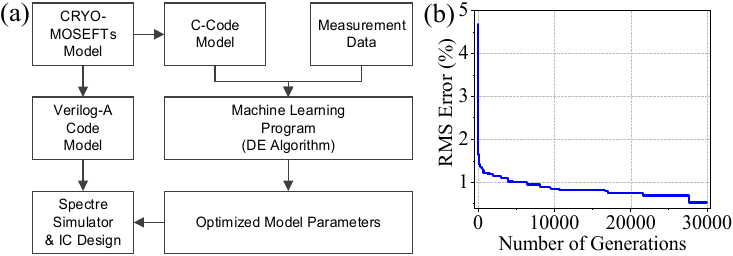}}
	\caption{(a) The flowchart of parameter extraction and modeling process of cryo-MOSFETs via a machine learning approach. (b) RMS error vs the number of generations in a certain optimization process. }
	\label{fig5}
\end{figure}

To obtain more accurate cryogenic MOSFETs (cryo-MOSFETs) model parameters, the parameters are optimized using a machine learning (ML) approach. As shown in Fig. \ref{fig5}(a), the cryo-MOSFETs model was written in both Verilog-A language and C language. The C language code is used to optimize the model parameters in the ML process. The differential evolution (DE) algorithm\cite{b13} is used in the optimization process because of its strong global convergence ability and high robustness. The measured data include three characteristics, i.e., the output characteristics and the transfer characteristics in both the linear and the saturation regions. The root-mean-square (RMS) error is introduced to evaluate the fitness of the optimized parameters, which is given by\cite{b7}: 
\begin{equation}
RMS error = \sqrt{\frac{1}{N} \times \sum_{i=1}^{n}(\frac{I_{measi}-I_{calci}}{I_{measimax}})^{2} } \times 100\%
\end{equation}
where N represents the number of data points, $I_{meas}$ and $I_{calc}$ are measured data and calculated data, respectively. $I_{measimax}$ is the maximum measured value. The total RMS error is obtained by averaging the RMS errors of the three characteristics above. As shown in Fig. \ref{fig5}(b), the RMS error decreases monotonically with the optimization process. Finally, the optimized parameters are input into the Verilog-A code model, the results of which can be calculated using commercial simulators and then used in IC design.
\subsection{Modeling}
\begin{figure*}[!t]
	\centering
	\includegraphics[scale=0.85]{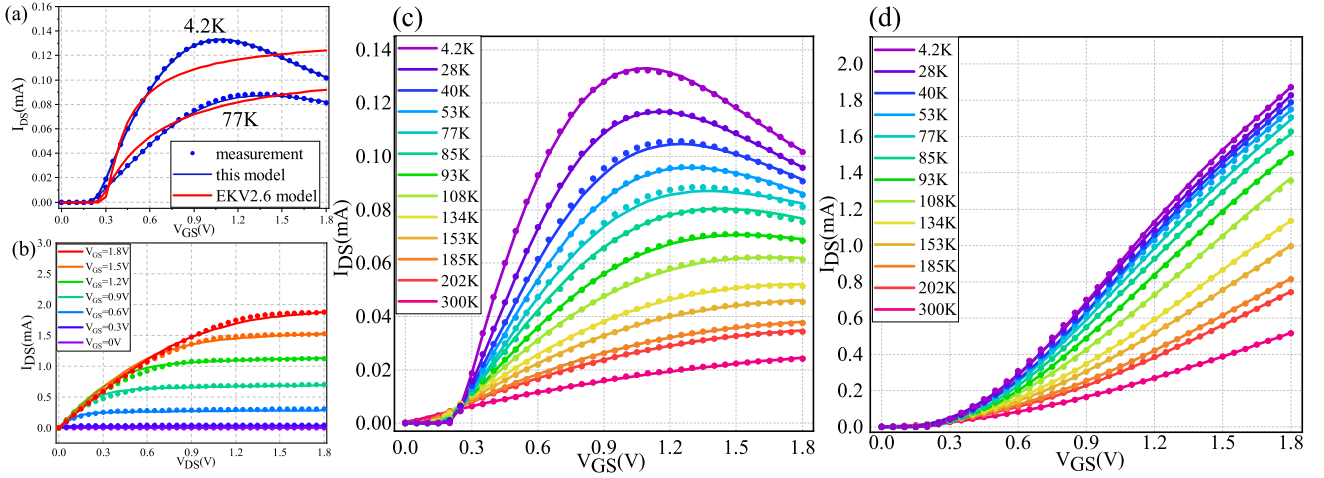}
	\caption{I-V curves of native MOSFETs with W/L$\,$=$\,$10$\,$$\mu m$/10$\,$$\mu m$ measured (symbol) and calculated (solid line) at various temperatures. (a) A comparison of the calculation results between this model and the EKV2.6 model at 77$\,$K and 4.2$\,$K. (b) Measurement and calculation results of the output characteristic at 4.2$\,$K. (c) Measurement and calculation results of the linear region transfer characteristic ($V_{DS}$$\,$=$\,$50$\,$mV) from 300$\,$K to 4.2$\,$K. (d) Measurement and calculation results of the saturation region transfer characteristic ($V_{DS}$$\,$=$\,$1.8$\,$V) from 300$\,$K to 4.2$\,$K.}
	\label{fig6}	
\end{figure*}
As shown in Fig. \ref{fig1}(b), a serious current drop occurs in the linear region transfer characteristics of native MOSFETs, which is because of the mobility behavior at low temperatures. Even after parameter optimization, the EKV2.6 model cannot describe this phenomenon accurately, as shown in Fig. \ref{fig6}(a). At room temperature, the mobility is mainly determined by the lattice (phonon) scattering and the surface roughness scattering. With increasing $V_{GS}$, $E_{eff}$ increases and the surface roughness scattering deteriorates, leading to reduced carrier mobility. 
\par At low temperatures, the average phonon number decreases rapidly and the phonon scattering is greatly reduced, even to the point of becoming negligible. The cryogenic mobility is determined by two main scattering processes, i.e., Coulomb scattering of the ionized impurities at lower $V_{GS}$ and surface roughness scattering at higher $V_{GS}$. The increase of $V_{GS}$ leads to an increase of electron concentration in the inversion layer and the Coulomb mobility is thus elevated. When $V_{GS}$ increases further, the Coulomb scattering becomes less effective than the surface roughness scattering, resulting in a reduction of the mobility. Therefore, the mobility shows a bell-shaped behavior at low temperatures\cite{b12}, which causes $I_{DS}$ to decrease after an initial increase in the linear region transfer characteristics. 
\par Because the EKV model is compact, concise, and accurate\cite{b7}, we selected the EKV2.6 model as the basis and then modified the mobility calculation equations to describe the DC characteristics of native MOSFETs accurately over the range from room temperature to cryogenic temperatures. By simplifying the theoretical calculation equations and analyzing the previous measurement results in \cite{b14,b15,b16,b17,b18}, we use the following formulas to describe the mobility of the Coulomb scattering ($\mu_{coul}$), the  surface roughness scattering ($\mu_{sr}$), and the phonon scattering ($\mu_{ph}$), respectively:
\begin{equation}
\mu_{coul}=\frac{A_{0}}{\left (1+q_{inv } / q_{0}\right)^{2}}
\end{equation}
\begin{equation}
\mu_{s r}=\frac{1}{A_{1} \times E_{eff }^{2}}\\
\end{equation}
\begin{equation}
\mu_{p h}=A_{2} \times\left(q_{inv}+q_{dep }\right)^{-1/3}\\
\end{equation}
where $q_{inv}$, $q_{dep}$ and $E_{eff}$ are the same physical quantities as (1), and $A_{0}$, $A_{1}$, $A_{2}$ are model parameters. $q_{0}$ is a parameter related to ionized impurities. $q_{inv}$ and $q_{dep}$ are calculated by the default equaptions of EKV2.6 model. According to Mathiessen's rule, the effective mobility ($\mu_{eff}$) is given by: 
\begin{equation}
\frac{1}{\mu_{\text{eff}}}=\frac{1}{\mu_{coul}}+\frac{1}{\mu_{s r}}+\frac{1}{\mu_{ph}}
\end{equation}
Eq. (7) is associated with EKV2.6 model by $\beta$$\,$=$\,$ $\mu_{eff}$$\times$$C_{OX}$$\times$$W/L$$\times$$10^{-4}$. $\beta$ is a default variable related to mobility in EKV2.6 model \cite{b19} and the unit of $\mu_{eff}$ is $cm^{2}/(V\cdot$s$)$. 
\begin{figure}[!t]
	\centerline{\includegraphics[width=\columnwidth]{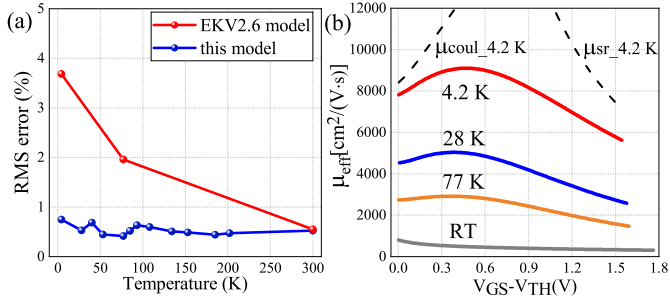}}
	\caption{(a) RMS error of this model and the EKV2.6 model at different temperatures. (b) Calculation results of the effective mobility at RT, 77 K, 28 K and 4.2 K, respectively. The dashed lines are $\mu_{coul}$ and $\mu_{sr}$ at 4.2 K, respectively. }
	\label{fig7}
\end{figure}
\par The calculation results are presented in Fig. \ref{fig6}. As expected, the results show the intersection behavior in the output characteristics at 4.2$\,$K, as shown in Fig. \ref{fig6}(b). Fig. \ref{fig6}(c) and (d) show the linear region and the saturation region transfer characteristics from 300$\,$K to 4.2$\,$K, respectively. 
The RMS error of the calculation results is less than 0.75$\%$ at each temperature, as shown in Fig. \ref{fig7}(a), and is much smaller than the corresponding results of the original EKV2.6 model. It is concluded that the proposed model is applicable from room temperature to liquid helium temperature.  
\begin{figure}[!t]
	\centerline{\includegraphics[width=\columnwidth]{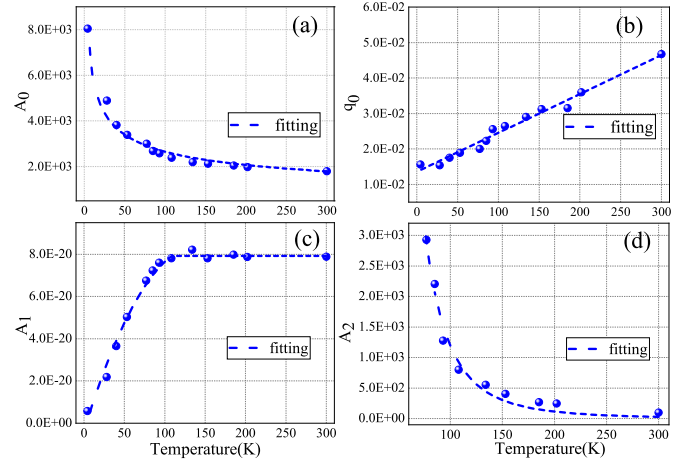}}
	\caption{The mobility-related parameters vs temperature. (a) $A_{0}$. (b) $q_{0}$. (c) $A_{1}$. (d) $A_{2}$. The points are the mobility-related parameters at specific temperatures extracted via the machine learning approach and the dotted lines are fitting lines.}
	\label{fig8}
\end{figure}
\par The variation of the parameters with temperature is also worthy of discussion. Because its influence is negligible, we did not consider the phonon scattering below 77$\,$K. Fig. \ref{fig8}(a) and (b) show the variations of Coulomb scattering related parameters (i.e., $A_{0}$, $q_{0}$). With decreasing temperature, the average velocity of the carriers increases, which means that carriers can pass through the Coulomb potential field of the ionized impurity atoms in a shorter time. Therefore, the carrier deflection angle under Coulomb scattering becomes smaller and the scattering probability is reduced, which results in a monotonic increase of $A_{0}$ with decreasing temperature. Additionally, the ionization rate decreases at low temperatures, and thus only a few impurities can be ionized to form Coulomb centers, resulting in the decrease of $q_{0}$. The combination of the two mechanisms above leads to a higher $\mu_{coul}$ value at lower temperatures. The variation of $A_{0}$ and $q_{0}$ with temperature can be fitted as follows, respectively: $A_{0}\!\!=\!\!1.4\times10^{4}\times T^{-0.357}$, $q_{0}\!\!=\!\!1.1\times10^{-4}T+0.014$. 
\par The variation of the surface roughness scattering coefficient ($A_{1}$) with temperature is shown in Fig. \ref{fig8}(c). $A_{1}$ remains almost constant above 110$\,$K but decreases with decreasing temperature below 110$\,$K, which is likely to be caused by the degeneracy effect. At cryogenic temperatures, the carriers become strongly degenerate\cite{b20}, resulting in a reduction of the surface roughness scattering under high electric fields\cite{b14}. The cryogenic degeneracy effect is  negligible above 110$\,$K, so $A_{1}$ is almost unchanged in that range. The variation of $A_{1}$ with temperature can be fitted as follows: $A_{1}\!=\!-4.4\times10^{-24}T^{2}+1.2\times10^{-21}T-3.5\times10^{-21}$ ($T<110\,K$) and $A_{1}\approx7.93\times10^{-20}$ ($T>110\,K$). Although $A_{1}$ is almost unchanged above 110$\,$K, due to $q_{inv}$$\,$$\approx$$\,$$Cox (V_{GS}-V_{TH})$, higher temperature leads to a smaller $V_{TH}$, i.e., $q_{inv}$$\,$$\,$ and$\,$$\,$$E_{eff}$$\,$$\,$increase, resulting in the decrease of $\mu_{sr}$ with increasing temperature. 
\par The variation of the phonon scattering coefficient is shown in Fig. \ref{fig8}(d). $A_{2}$ increases significantly with decreasing temperature and can be fitted as $A_{2}=6.6\times10^{9}T^{-3.4}$. Furthermore, the calculation results of $\mu_{eff}$ at specific temperatures are shown in Fig. \ref{fig7}(b). The variation trend of $\mu_{eff}$ shows a bell-shaped behavior at low temperatures, which is consistent with the theoretical predictions. 

\section{Conclusion}
Native MOSFETs can be used in high-speed or low power consumption circuits. In this paper, commercial 0.18$\,$$\mu$m native MOSFETs are characterized in the temperature range from 300$\,$K to 4.2$\,$K. With decreasing temperature, $V_{TH}$ increases up to $\sim$0.25$\,$V (W/L$\,$=$\,$10$\,$$\mu$m/10$\,$$\mu$m) and the improved $SS$$\,$$\approx$$\,$14.30$\,$mV/dec$\,$@$\,$4.2$\,$K. The $I_{off}$, the GIDL induced leakage, and the ON/OFF ratio are ameliorated obviously, but the DIBL effect shows deterioration. In addition, the step-up effect and the $G_{m}$-peak effect are observed and discussed. By using Coulomb scattering, surface roughness scattering, and phonon scattering to calculate the carrier mobility, a compact model of large size native MOSFETs suitable for the range of 300$\,$K to 4.2$\,$K is proposed. This model is able to fit the DC characteristics very well, particularly the current drop$\,$$\,$behavior in the linear region. In addition, the mobility-related parameters are extracted via a machine learning approach and the temperature dependences of the scattering mechanisms are analyzed. This work contributes to the research of cryo-MOSFETs physics and also lays a foundation for cryo-CMOS circuit design.



\end{document}